\documentclass[twocolumn]{aastex63}

\received{\today}
\usepackage{graphics}
\usepackage{amsmath,amsfonts,amssymb}
\usepackage[T1]{fontenc}
\usepackage[utf8]{inputenc}
\usepackage{epstopdf}
\usepackage{babel}
\submitjournal{ApJ}

\shorttitle{Viewing angle effects in quasar application to cosmology}
\shortauthors{Prince et al.}
\graphicspath{{./}{Figure/}}

\begin{document}

\title{Viewing angle effects in quasar application to cosmology}


\author[0000-0002-1173-7310]{Raj Prince}
\email{raj@cft.edu.pl}
\author[0000-0001-5848-4333]{ Bo\.zena Czerny}
\email{bcz@cft.edu.pl}
\affiliation{Center for Theoretical Physics, Polish Academy of Sciences, Al.Lotnikow 32/46, 02-668, Warsaw, Poland}
\author{Agnieszka Pollo}
\affiliation{National Centre for Nuclear Research, ul. Pasteura 7, 02-093 Warsaw, Poland}



\begin{abstract}
The symmetry axes of active galactic nuclei (AGN) are randomly distributed in space but highly inclined sources are heavily obscured and are not seen as quasars with broad emission lines. The obscuring torus geometry determines the average viewing angle, and if the torus geometry changes with the redshift, this average viewing angle will also change. Thus the ratio between the isotropic luminosity and observed luminosity may change systematically with redshift. Therefore, if we use quasars to measure the luminosity distance by evaluating the isotropic absolute luminosity and measuring the observed flux, we can have a redshift-dependent bias which can propagate to cosmological parameters. We propose a toy model for testing the effect of viewing angle uncertainty on measurement of the luminosity distance. The model is based on analytical description of the obscuring torus applied to one-parameter observational data. It illustrates the possible change of the torus covering factor between the two chosen redshift ranges.  We have estimated the possible error on specific cosmological parameters (H$_0$, $\Omega_m$) for the flat $\Lambda$CDM cosmology if a method is calibrated at low redshift and applied to the higher redshift. The errors on cosmological parameters due to potential dependence of viewing angle on redshift are found to be potentially significant, and the effect will have to be accommodated in the future in all quasar-based cosmological methods.  A careful systematic study of AGN mean viewing angle across redshift is necessary, with the use of appropriate samples and models which uniquely determine the inclination of each source.

\end{abstract}

\keywords{cosmology: distance scale -- galaxies: active -- galaxies: general -- quasars: general}

\section{Introduction}

Quantifying the accelerated expansion of the Universe is one of the key issues of the cosmology \citep[for a recent reviews, see e.g.][]{matarrese2011,rubin2016,czerny2018}. Various probes are used to this purpose, like observations of the Cosmic Microwave Background (CMB; \citealt{Planck_2018}), Supernovae Ia (SN Ia; \citealt{Riess_1998}, \citealt{Perlmutter_1999}), Baryon Acoustic Oscillations (BAO; \citealt{bassett_2010}), gravitational lensing (\citealt{Wong_2020}), or gamma-ray bursts (\citealt{Schaefer_2003}). Quasars, or more generally, Active Galactic Nuclei (AGN) also joined the class of sources with cosmological applications, and several specific methods to use these objects were proposed: continuum time delays \citep{collier1999,cackett2007}, emission line time delays \citep{watson2011,haas2011,czerny2013}, broad-band spectral shape (nonlinear dependence between the UV and X-ray flux ,\citep{risaliti2015,lusso2019}), virial broadening estimator studies \citep{Franca_2014,marziani2014,marziani2020}, statistical properties of the quasar distribution reflecting the large scale structure \citep[e.g.][]{decarvalho2020}, and quasar strong lensing \citep{wong_treu2020}. 

Several recent measurements based on different methods imply the tension between the Hubble constant $H_0$ determination based on early Universe and the value coming from the relatively local measurements \citep[see e.g.][for recent reviews]{riess_review2019,verde2019}. However, high accuracy of measurements is necessary, and the control of the systematic errors is critical, since the systematic error of 0.2 mag in the luminosity distance is enough to eliminate the $H_0$ tension. 

The use of quasars for cosmology has important advantages. Quasars are numerous, and they nicely cover the broad range of redshifts, starting from relatively nearby AGN to redshifts of order of 7 (\citealt{Mortlock_2011}, \citealt{Banados_2018}). They also do not show strong extinction effects, since bright quasars efficiently clear the line of sight through the host galaxy, and in addition the color selection allows to remove the highly reddened minority from the sample. They also do not seem to show evolutionary effects related to the metallicity since even high redshift quasars are metal-rich, with solar or even super-solar abundance.

Quasar-based constraints of the cosmological parameters were already very successful. Combining the quasar method based on the broad-band spectral shape with several other cosmological probes \citep{lusso2019} showed that there is a 4$\sigma$ tension with the standard $\Lambda CDM$ cosmology. 
Quasar-based results will flourish in the next years, with more AGN being a subject of systematic reverberation monitoring \citep{dupu2015,grier2017,dupu2018,grier2019,homayouni2020,cackett2020}, and with the future surveys approaching, like LSST \citep{ivezic2019}.

Since the number of sources is growing, the cosmological constraints will be determined with smaller and smaller statistical errors. However, the systematic errors inherent in the methods will not disappear with the rise of the statistical samples. In the case of AGN, one of such potentially important issues is the effect of the viewing angle of an active nucleus. AGN are not point-like sources, they show significant dependence of their appearance on the inclination angle of an observer with respect to the symmetry axis which underlies the AGN classification \citep[see e.g.][for a review] {Antonucci_1993, Urry_1995, Netzer_2015}. This by itself does not pose a problem as long as, statistically, there is no trend in the average viewing angle with the redshift.   However, if such a trend is likely to exist, it can be expected to affect the conclusions about the redshift dependence of the Universe expansion rate. The aim of the present paper is to check if such trends are likely, and if so, whether they may lead to biased estimation of cosmological parameters. A hint that it may happen is provided by 
\citet{Gu_2013} who has found the redshift evolution in the covering factor of the torus. We use this data in our experiment. 

The paper is organised as follows.
In section 2, we discuss the method, where we outline our approach to the effect of viewing angle on measurement of the disk luminosity, followed by geometry description, analytical description of the disk/torus system, selection of the observational data, and predicted effect on cosmological parameters  are given in section 2.1, 2.2, 2.3 and 2.4, respectively. In sections 3 $\&$ 4, we present the discussion in the context of quasar cosmology.

\section{\bf Method}
AGN is believed to have a supermassive black hole at the center which accretes the matter from the surrounding. In vigorously accreting systems this matter forms cold Keplerian accretion disks. Close to the disk, the broad line region forms. This central part of an AGN is usually surrounded by a dusty/molecular torus. 
Thus the viewing angle plays an important role in how the AGN appears, and consequently, in how it is classified. The disk is flat, so the observed disk flux decreases with an increase of the viewing angle. In addition, the observer looking at very high viewing angle (measured from the symmetry axis) will not be able to see the direct disk emission, and instead will register only a processed emission from the molecular torus (plus, eventually, a scattered emission). At intermediate viewing angles both the disk anisotropy and the partial obscuration are important. Therefore, the observed radiation from any AGN mostly depends on two important parameters e.g. viewing angle ($i$) and the optical depth ($\tau$) along the line of sight. Here
we consider our study as a toy model illustrating potential dangers of negligence of all viewing angle issues in quasar cosmology.\\

\subsection{\bf Description of the Geometry} 
\label{subsec:floats}

\begin{figure}
    \centering
    \includegraphics[scale=0.6]{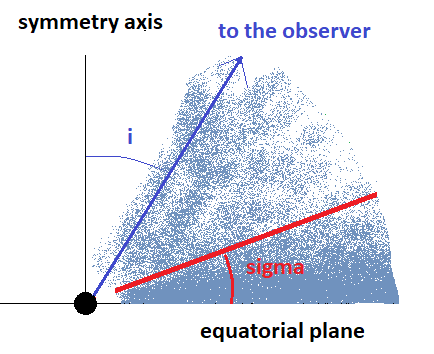}
    \caption{Cartoons for a dusty torus with half opening angle $\sigma$ and with decreasing density away from the equatorial plane. The viewing angle is defined as $i$ measured from the symmetry axis.} 
    \label{fig:my_label}
\end{figure}

The AGN geometry to be considered in this work is similar to what is given by \citet{Elitzur_2008}, and shown in Figure 1, where torus is considered to have a smooth density distribution of dust in angular direction. 
The inclination angle measured from symmetry axis is denoted as $i$. 

We assume that the radiation produced in the disk is modified by the torus before it reaches the observer, and the effect depends on the viewing angle. Let's assume that the $\tau$ is the optical depth of molecular torus and it is defined as,
\begin{equation}
\label{eq:tau}
\tau = A e^{-(\beta/\sigma)^2}
\end{equation}
where 'A' is the normalization constant, $\beta$ (=$\pi$/2-$i$) is the angle measured from the equatorial plane, and  $\sigma$ is the characteristic width of the torus. We thus use just the projection of the torus onto the sky plane of the central source and neglect possible variations in the radial distribution of the dusty material. Therefore, constants $A$ and $\sigma$ are the only two free parameters of the torus model itself. The third parameter is the viewing angle.

\subsection{\bf Disk/torus system }
\label{sect:model}
Here we consider a toy model which allows us to see how the 
change in the covering factor measured by \citet{Gu_2013} can relate to the torus parameters and the viewing angle, and further, how it can affect the measured luminosity in case of high redshift AGN (2.0 $\leq z \leq2.4$) if we fix the geometry by analyzing the low redshift sample (0.7 $\leq z \leq1.1$). We do not aim here at actual measurement of the observational trend, as this will have to be done in the future with better torus model and a large sample of objects over a redshift range with broad band spectra which will allow to measure uniquely the model parameters, including the viewing angle, from the data.

The disk luminosity seen by the observer is anisotropic since the AGN standard accretion disk is geometrically thin so the disk flux is proportional to the $cosine$ of the viewing angle ($i$), i.e. the intrinsic disk flux $L_{disk}$ seen by observer in the absence of the dusty torus is:
\begin{equation}
\label{eq:disk}
L_{disk}(i) = 2 L^{aver}_{disk} \cos i, 
\end{equation}
where L$^{aver}_{disk}$ is the intrinsic average disk luminosity. When the observer is looking along the symmetry axis, where $i$ = 0, the observed disk luminosity, L$_{disk}$, would be 2L$^{aver}_{disk}$ \citep[see e.g.][for a reference]{frank2002}. This anisotropy effect has been neglected in \citet{Elitzur_2008} as well as in \citet{Gu_2013} analysis. In principle, additional factor related to the limb-brightening could have been also included here but for the sake of simplicity we neglect it in the current approach.  

Finally, the disk intrinsic anisotropic emision is modified in a viewing-angle dependent way when the light is passing through the dusty torus which gives us the final expression for the observed disk flux:
\begin{equation}
L_{disk}(i) = 2 L^{aver}_{disk} \cos i ~e^{-\tau}, 
\end{equation}
where $\tau$ is the optical depth of dusty torus.
The IR radiation produced in the molecular torus comes from the nuclear emission which is intercepted by the torus. In order to calculate it from the model, assuming that the intrinsic disk emission and torus parameters are known, we perform an integral over all viewing angles: 
 \begin{equation}
L_{IR} = 2 L_{disk}^{aver} \int_{0}^{\pi/2}\sin \beta * cos\beta *(1 - e^{-A e^{-\beta^2 / \sigma^2}})*d \beta.
\end{equation}
Here we use the angle $\beta$ measured from the equatorial plane.

We further assume that the emission of the torus in the IR band is isotropic. This is an oversimplification in comparison with the approach of \citet{Elitzur_2008} based on radiative transfer, but we do not aim to study the spectral features in the IR band. This assumption of isotropy is relatively correct if the torus is optically thin and/or we observe the emission at relatively longer wavelengths. Here we concentrate on a simple approach to overall radiation budget.

Thus we define the covering factor (CF) of the AGN  as the ratio of observed IR to the intrinsic disk luminosity: 
\begin{equation}
 CF = L_{IR}/L_{disk}^{aver} = L_{IR}/L_{disk}*L_{disk}/L_{disk}^{aver}.
\end{equation}
 
 On the other hand, if we assume that the IR emission is isotropic but the disk emission is highly anisotropic, then what we measure directly from the data is instead the ratio:
 \begin{equation}
 \label{eq:rat1}
  L_{IR}/L_{disk} = \frac{L_{IR}/L_{disk}^{aver}}{L_{disk}/L_{disk}^{aver}}.
 \end{equation}
 
 The numerator in the above equation is estimated through the numerical integration from equation (4). The denominator is the re-written form of equation (3),
 \begin{equation}
L_{disk}/L^{aver}_{disk} = 2  \cos i *e^{-\tau}, 
\end{equation}
after substituting the value $i$ = $\pi$/2 - $\beta$, the equation (7) transforms to,
\begin{equation}
\label{eq:disk_norm}
L_{disk}/L^{aver}_{disk} = 2  \sin \beta *e^{-A*e^{-(\beta / \sigma)^2}}.
\end{equation}

Although the viewing angle has such an importance for the appearance of AGN, we rarely measure it for individual sources. A range of viewing angles would lead to a scatter in the ratio of $L_{disk}/L^{aver}_{disk}$. This scatter is around some mean value, unknown a priori and it needs to be determined observationally. This value is sensitive to the properties of the torus. 

In reverberation mapping we use only type 1 AGN, with emission lines well visible. The torus shields the BLR from the observer for high inclinations for which the torus becomes optically thick. The range of the viewing angles shielding the nucleus depends on the dust spatial distribution, i.e. dust properties. In addition, the transition from unshielded to shielded case is continuous for most torus models. Here we assume that the line of sight with the value  $\tau$ = 1 separates the type 1 and type 2 AGN. We mark the corresponding value of the angle $\beta$ as $\beta_0$, and we use Equation~\ref{eq:tau} to estimate its value knowing the torus parameters. Our torus model gives 
\begin{equation}
\label{eq:beta_0}
\beta_0 = \sigma \sqrt{ln(A)}
\end{equation}
and hence $\sin \beta_0$ = $\sin$($\sigma$ $\sqrt{ln(A)}$).

Thus type 1 sources are those with the viewing angle between 
 $\beta$ = 0 to $\beta$ = $\beta_0$, and statistically the mean (effective) viewing angle is given by $\sin \beta_{eff}$ = 1/2(1+ $\sin \beta_0$) which, by substituting the value of $\beta_0$, reduces to 1/2(1+ $\sin$($\sigma$ $\sqrt{ln(A)}$)).

Therefore, if we consider a sample of randomly oriented type 1 sources, we can substitute $\beta$ with  $\beta_{eff}$ in Equation~\ref{eq:disk_norm}, and use Equation~\ref{eq:rat1} to get the average value for the normalized disk luminosity,
\begin{equation}
\label{eq:basic_ratio}
  L_{IR}/L_{disk} = \frac{L_{IR}/L_{disk}^{aver}}{(1+\sin (\sigma \sqrt{ln(A)})) *e^{-A*e^{-(\beta_0 / 2\sigma)^2}}}.
 \end{equation}
Values of A and $\sigma$ are model parameters, $\sigma$ can vary between 0 to 90 degree, but A must be larger than 1 since otherwise we would have no type 2 sources and the definition of $\beta_0$ in Equation~\ref{eq:beta_0} would not apply. 
If the ratio of $L_{IR}/L_{disk}$ is measured observationally we can constrain the geometry of the torus, and in particular the average viewing angle and intrinsic extinction.

\subsection{\bf Observational data} \label{subsec:tables}
\citet{Gu_2013} has studied the sample of quasars from DR9Q \citep{Paris_12} and DR7Q \citep{Schneider_10} catalog by combining the photometric data from SDSS with the data from other observatories, including WISE \citep{Wright_10}, UKIDSS \citep{Lawrence_07}, and GALEX \citep{Martin_2005}. 

The distribution of quasars sample in DR9Q and DR7Q peaks in two different redshift range. The distribution with redshift can be seen in \citet{Paris_12} and \citet{Schneider_10}, where large number of sources in DR9Q found in redshift range 2.0-2.4 with median 2.2 and in DR7Q the source distribution with redshift shows two peaks, one at 0.8 and another at 1.6. Now to consider the significantly different redshift range for low and high redshift sources \citet{Gu_2013} focused on the peak 0.8 and selected the sources within the redshift range 0.7-1.1.
The multi-wavelength data collected from other telescopes like WISE, GALEX, UKIDSS along with SDSS are used to constrain the sample size and finally the multi-wavelength SED was produced. The SED were used to estimate the IR and bolometric luminosity for all the sources and further, their ratio is defined as covering factor (CF) by \citet{Gu_2013}. The distribution of sources for low and high redshift with CF are shown in Figure 5 of \citet{Gu_2013}, and their corresponding median values differ significantly, suggesting a clear evolution with redshift. They have done the Kolmogorov-Smirnov statistic (KS) test to confirm the difference between low and high redshift CF distribution. Both sub-samples show a wide distribution of the values but there is an overall strong shift in the reported values towards higher dust covering at large redshift despite no clear effect in distributions of black hole mass and Eddington ratio. 

It can always be argued that the difference seen in the torus geometry at two different redshift ranges possibly results from the biases present in the sample selection. But it can only be verified in future by having more observed sources over a larger redshift range, with more precise observations. 

For the use in the present paper we concentrate only on the median values of the IR to bolometric luminosity (CF), which is 0.478 for the low redshift subsample and 0.871 for the high redshift subsample. In their paper, they report directly the ratio of the IR to the bolometric luminosity, and they consider that as a measurement of the covering factor (CF) assuming a spherically symmetric (isotropic) emission of the central source and the even simpler torus model without stratification along the $\beta$ angle. In our approach, their measurement actually corresponds to the ratio defined by Equation~\ref{eq:basic_ratio}.

\begin{table}
\centering
\begin{tabular}{c c}
mean~redshift & CF     \\
\hline
\hline
0.8 & 0.478 $\pm$ 0.01  \\
2.2 & 0.871 $\pm$ 0.03  \\
\end{tabular}
\caption{Summary of the parameters from \citet{Gu_2013} used in the present paper.}
\label{tab:redshift}
\end{table}

The change of this factor between the low and the high redshift sub-sample implies the systematic change of the effective viewing angle in quasars with redshift. If this effect is ignored in the studies using quasars for cosmology this will lead to clear systematic apparent drift in cosmological parameters. In the next section, we will describe how the negligence of this information affects the results. We summarize the parameters adopted from \citet{Gu_2013} in Table \ref{tab:redshift}.

Since \citet{Gu_2013} provides a single value for each source, we cannot uniquely derive our two torus parameters and the viewing angle from their measurements. The assumption that we concentrate on the mean viewing angle relates the viewing angle to the torus parameters, but still two parameters remain to be determined from a single measurement. Therefore, we cannot expect to determine the actual trend with the redshift, ready for cosmological applications, but nevertheless we obtain a very interesting illustration of what {\it may} happen if the issue is neglected.

\subsection{\bf Application to Cosmology}

We first provide simplified analytical estimates how the lack of knowledge of the viewing angle of a quasar leads to incorrect determination of the Hubble constant from the reverberation mapping of a single quasar. Next we use the data from Table \ref{tab:redshift} to show how large errors for the estimate of the cosmological parameters can result if the systematic change in the torus parameter between the lower redshift and higher redshift sub-samples is ignored.

\subsubsection{\bf Simple example of viewing angle role applicable at very low redshift}
\label{sect:very_simple}
In standard cosmological approach to use quasar reverberation measurements, we neglect the issue of the viewing angles since we do not measure them routinely for the reverberation measured sources.
We then measure the time delay for a given source, $\tau_d$, and we measure the observed monochromatic flux, for example at 5100 $\AA$, F$_{obs}$. 
We use the theoretical formula, neglecting the issue of the viewing angle, to define the observed monochromatic luminosity,
\begin{equation}
\log L_{44,5100} = -2.61 + 2 log (\tau_d),
\end{equation}
where the constant was calculated assuming the dust temperature T$_{dust}$ = 1000 K, and effective viewing angle, i$_{eff}$ = 39.2$^{\circ}$ in Equations (2), (3), and (4) of \citet{Czerny_2011}. Here the time delay $\tau_d$ of the H$\beta$ line with respect to the continuum at 5100 \AA ~, is measured in days, and the monochromatic luminosity at 5100 \AA~ in units of $10^{44}$ erg s$^{-1}$ cm$^{-2}$. Thus, the measurement of the time delay gives us the absolute luminosity of the source. 
Knowing the absolute luminosity one can estimate the 
source luminosity distance by following the relation,
\begin{equation}
D_L^2 = {L_{5100} \over 4 \pi F_{5100}},
\end{equation} 
where $F_{5100}$ is the observed flux measured at 5100 \AA, easily available for the studies quasar. Derived $D_L$ can be compared to the cosmological model through the prescription for the cosmological distance $D_L(z,H_0,\Omega_{\Lambda},\Omega_m)$ if we know the redshift, $z$ (also easily measured for the quasar from the position of the emission line). Thus, measuring $\tau_d$, $z$ and $F_{5100}$, we can do cosmology. In particular, if the redshift is much smaller than 1, we can use a linear part of the Hubble flow determine the Hubble constant:
\begin{equation}
H_0 = const {z F_{5100}^{1/2} \over \tau_d}.
\end{equation}
 
The problem appears if the viewing angle is not consistent with the adopted value $i_{eff} = 39.2^{\circ}$ adopted by \citet{Czerny_2011}. So one can ask the question, what happens if we use a single object to obtain $H_0$ but the source is seen at an angle $i$ different from $i_{eff}$ ?

Now we neglect the effect of the viewing angle in time delay geometrical setup \citep[see Equation~3 of][]{Czerny_2011} since this effect is highly dependent on the BLR 3-D geometry. In that case the time delay measurement give us the true viewing-angle independent absolute luminosity (and the product of the black hole mass and accretion rate), where the constant was adjusted to the viewing angle $39.2^{\circ}$. However, the {\it measured} flux, $F_{5100}$ would still depend on the actual viewing angle, so the luminosity distance would contain the term $({1 \over cos i})^{-1/2}$, and the measured Hubble constant would actually be 
\begin{equation}
H_0 = const \left({\cos i \over \cos i_{eff}}\right)^{1/2}{z F_{5100}^{1/2} \over \tau_d}.
\end{equation}
This means we will be making an error by a factor $(cos i/cos i_{eff})^{1/2}$ in H$_0$ if this factor is ignored, and that is mostly the case since we did not measure the viewing angle to the source. This is the problem in any quasar reverberation mapping, either based on emission line delay, or continuum time delay (see Equation~15 of \citealt{cackett2007}). In addition, in this simplified approach we also neglected the effect of the extinction due to the dusty torus.

Through this article, we tried to shed some light on the issue of possible systematic errors due to disk emission anisotropy and the dust properties, which might affect the quasar application to cosmology.

\subsubsection{\bf Error estimate based on the data of \citet{Gu_2013}}

Now we use the full model as described in Section~\ref{sect:model}, we include the disk anisotropy as well as the extinction caused by the torus, and we do not limit ourselves to very low redshift, so the approach is not analytical till the very end, as in Section~\ref{sect:very_simple}. However, our approach still remains relatively simple. 

We now assume that all the parameters for the low redshift (0.7 $\leq z \leq$ 1.1) subsample are properly adjusted to the correct cosmological model, and we study the errors which would be made if the cosmological parameters are estimated from the large redshift (2.0 $\leq z \leq$ 2.4) subsample but the change in the torus properties visible in the data of \citet{Gu_2013} is neglected.

When we constrain the cosmology using the high redshift quasar sample, we measure the luminosity distance and we should use the geometry appropriate for the high redshift sources. Considering all the appropriate effect, the observed flux can be defined as
\begin{equation}
F_{5100} \propto \left(\frac{\tau_d^2 \cos i~ e^{-\tau}} {D_L^2} \right) 
\end{equation}
where $\tau$ is the optical depth measured along the line of sight and $\tau_d$ is the time delay. The above equation is similar to the equation defined by \citet{cackett2007} in equation (14), though they have not considered the effect of optical depth. The luminosity distance from the above equation can be defined as,
\begin{equation}
D_L^2 \propto \left(\frac{\tau_d^2 \cos i~ e^{-\tau}} {F_{5100}} \right) 
\end{equation}
or
\begin{equation}
D_L \propto {\tau_d (\cos i)^{1/2} e^{-{\tau}/2} \over (F_{5100})^{1/2}}.
\end{equation}
As we noted before, we aim at using a statistical sample so we not need to know the viewing angle of the individual sources. However, we need to know the mean, or effective viewing angle in each redshift bin. Thus, if one wants to estimate the correct luminosity distance at high redshift bin, one should use the effective viewing angle at high redshift (cos$i_{eff}$(high-z)).  If the effective viewing angle appropriate for low redshift bin is used, then the luminosity distance at high redshift bin measured by using the low redshift geometry would be incorrect. The correct and incorrect luminosity distance at high redshift bin can be defined as,
\begin{equation}
D_L^{correct} \propto \left( \tau_d (\cos i_{eff}(high-z))^{1/2} e^{-{\tau_{high}}/2} \over (F_{5100})^{1/2} \right)
\end{equation}
and
\begin{equation}
D_L^{incorrect} \propto \left( \tau_d (\cos i_{eff}(low-z))^{1/2} e^{-{\tau_{low}}/2} \over (F_{5100})^{1/2} \right)
\end{equation}
The relative error in measuring the luminosity distance caused by the incorrect use of the low redshift geometry to high redshift sample  can be estimated as by taking the ratio of equation (19) to (18),
\begin{equation}
{D_L^{incorrect} \over D_L^{correct}} = \left(
\cos i_{eff}(low-z) \over \cos i_{eff}(high-z) \right)^{1/2}{e^{-\tau_{low/2}} \over e^{-\tau_{high/2}}}
\label{eq:DL_rat}
\end{equation}
where the time delay $\tau_d$ does not appear in the final equation (20). 

\begin{figure}
    \centering
    \includegraphics[scale=0.52]{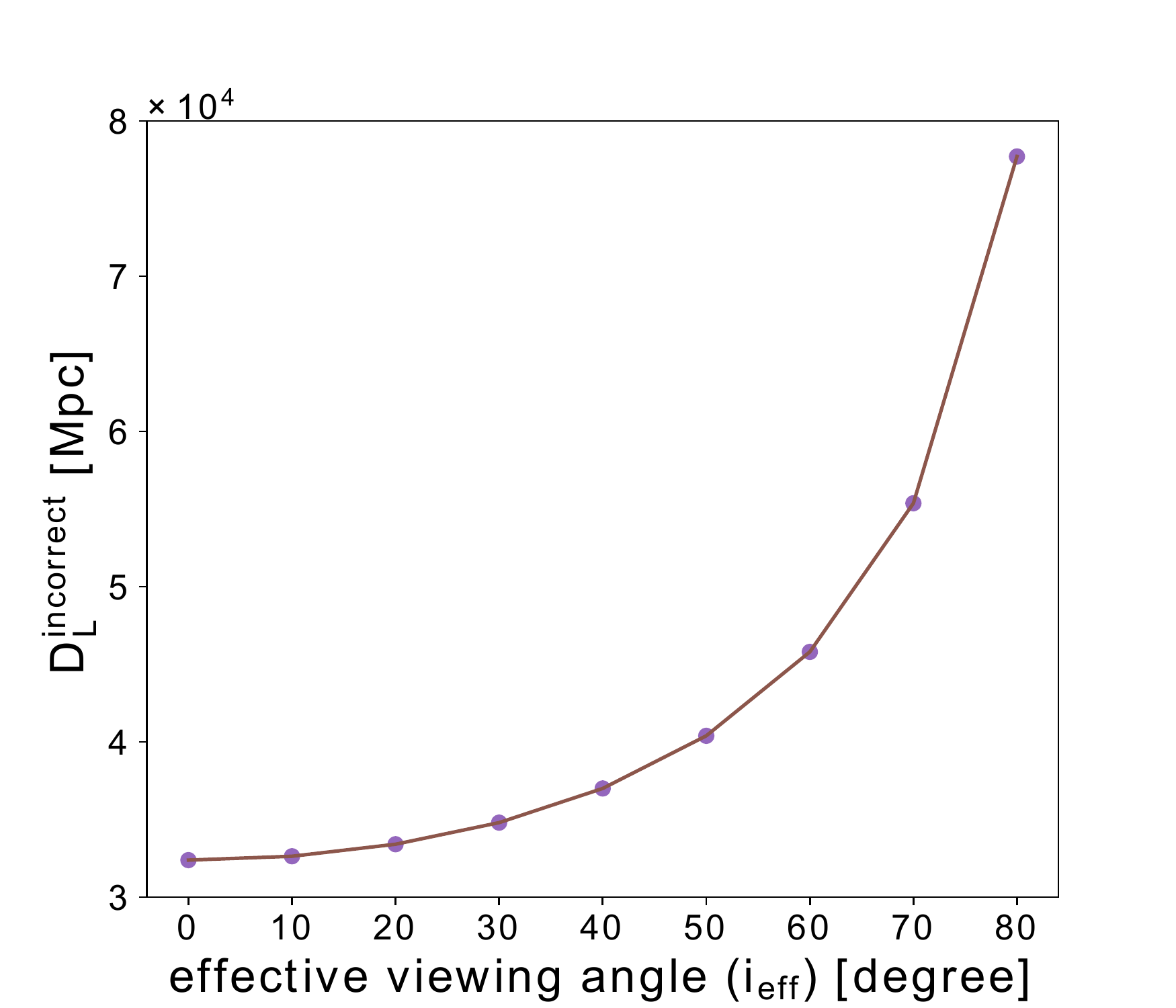}
    \caption{The variation of Luminosity distance for high redshift AGN with their effective viewing angle from equation(20). }
    \label{fig:DL_i}
\end{figure}

In Figure~\ref{fig:DL_i}, we show the variation of luminosity distance for high redshift objects (assuming at lower redshift all the measurements are correct) from equation (20). The luminosity distance at lower redshift is estimated for standard cosmology (\citealt{Planck_2018}) and the effective viewing angle, i$_{eff}$, is set at 39.2$^{\circ}$. In addition, optical depth is assumed to be constant for the purpose of this particular plot and the viewing angle is chosen within the allowed range. 

The error in the luminosity distance can be propagated to an error in the Hubble constant, $H_0$, and can be represented by following relation, if we assume that we know the other cosmological parameters,
\begin{equation}
D_L^{incorrect} = D_L(H_0^{incorrect},\Omega_m,\Omega_{\Lambda}).
\end{equation}
To reduce the number of free parameters, we consider the flat cosmology, where $\Omega_m$ + $\Omega_{\Lambda}$ = 1, and the equation (21) is finally modified to,
\begin{equation}
    D_L^{incorrect} = D_L(H_0^{incorrect},\Omega_m^{correct})
    \label{eq:H0_incorrect}
\end{equation}
where $\Omega_m^{correct}$ is the value taken from \citet{Planck_2018}.
The equation (22) can only be solved numerically. We have used the correct value of H$_{0}$ and other cosmological parameters from the \citet{Planck_2018} to estimate the correct luminosity distance. Using the correct luminosity distance in the equation (20) we estimated the incorrect luminosity distance at higher redshift cause by the geometry mismatch. And again we have used the equation (22) to estimate the H$_0^{incorrect}$ for the incorrect luminosity distance at fixed value of other cosmological parameters. 

Similarly, we can assume that the Hubble constant is known from other measurements but the aim is to obtain the $\Omega_m$. In this case we have to solve the equation
\begin{equation}
    D_L^{incorrect} = D_L(H_0^{correct},\Omega_m^{incorrect})
    \label{eq:Omega_m_incorrect}
\end{equation}
in order to see the error caused by inappropriate assumption about the torus geometry for high redshift sample.

To perform this exercise, we need the observational constraints for the mean viewing angles and optical depths at low and high redshift samples.




\section{\bf Results}

\subsection{\bf Constraints for the torus evolution with redshift and the corresponding change of the viewing from the Minfeng Gu's sample} \label{subsec:autonumber}

In section 2, we have discussed our simple torus model which we use to estimate the CF analytically. The equation (10) is the final equation which gives the CF for different values of 'A' and $\sigma$. By comparing our analytical value of CF to the observational values from \citet{Gu_2013}, we formulate constraints for the two torus parameters. Since, at a given redshift, we measure only one parameter (see Table~\ref{tab:redshift}) while the model has two parameters, the results are not unique.

\begin{figure}[ht!]
    \centering
    \includegraphics[scale=0.45]{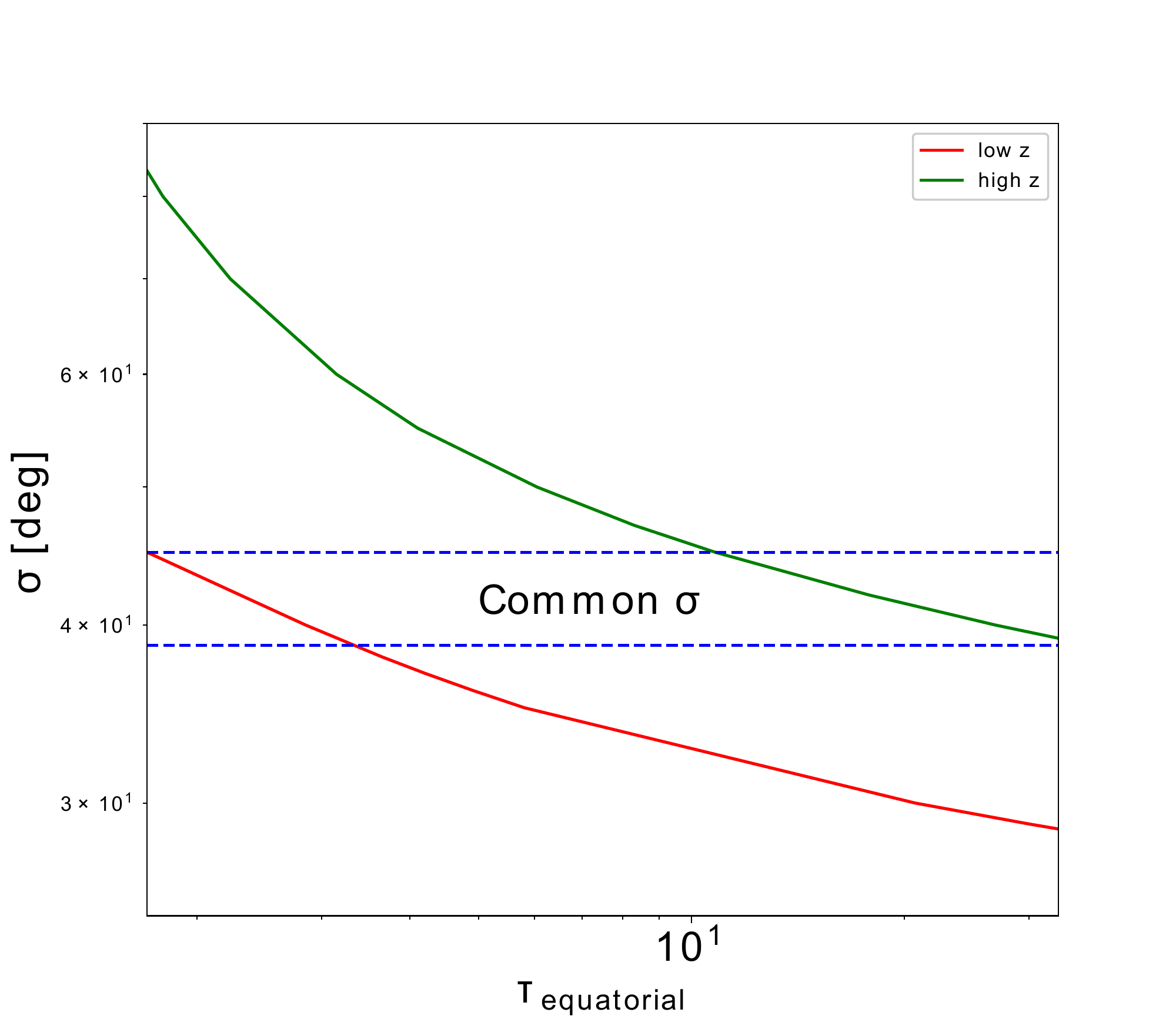}
    \caption{The relation between the torus parameters for low and high redshift sources based on constraints coming from the data in Table~\ref{tab:redshift}} 
    \label{fig:general}
\end{figure}

\begin{figure}
    \centering
    \includegraphics[scale=0.55]{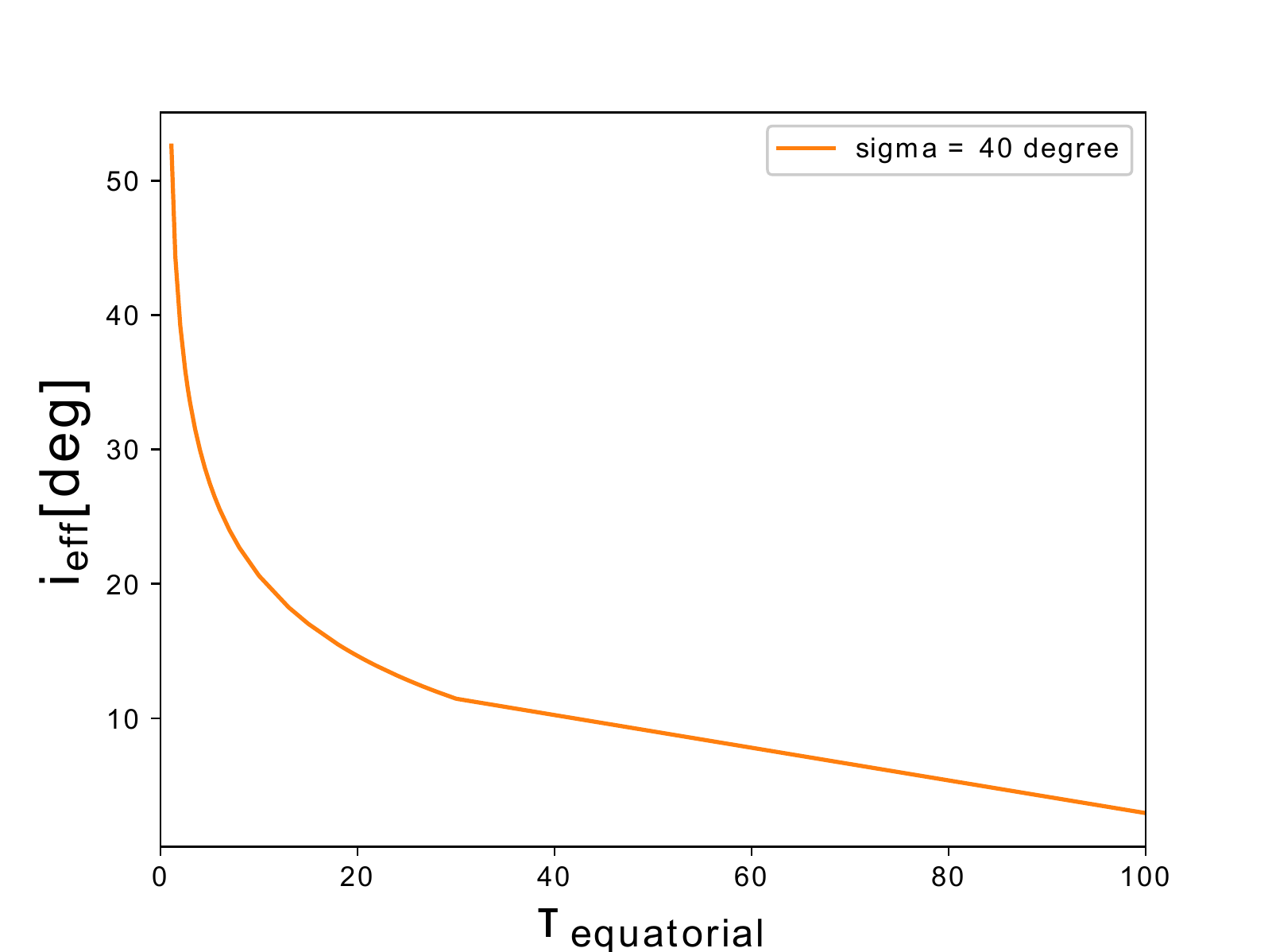}
    \caption{The effective viewing angle for the torus with $\sigma = 40^{\circ} $ as a function of torus optical depth.}
    \label{fig:i_eff_fixed_sigma}
\end{figure}

\begin{figure}
    \centering
    \includegraphics[scale=0.55]{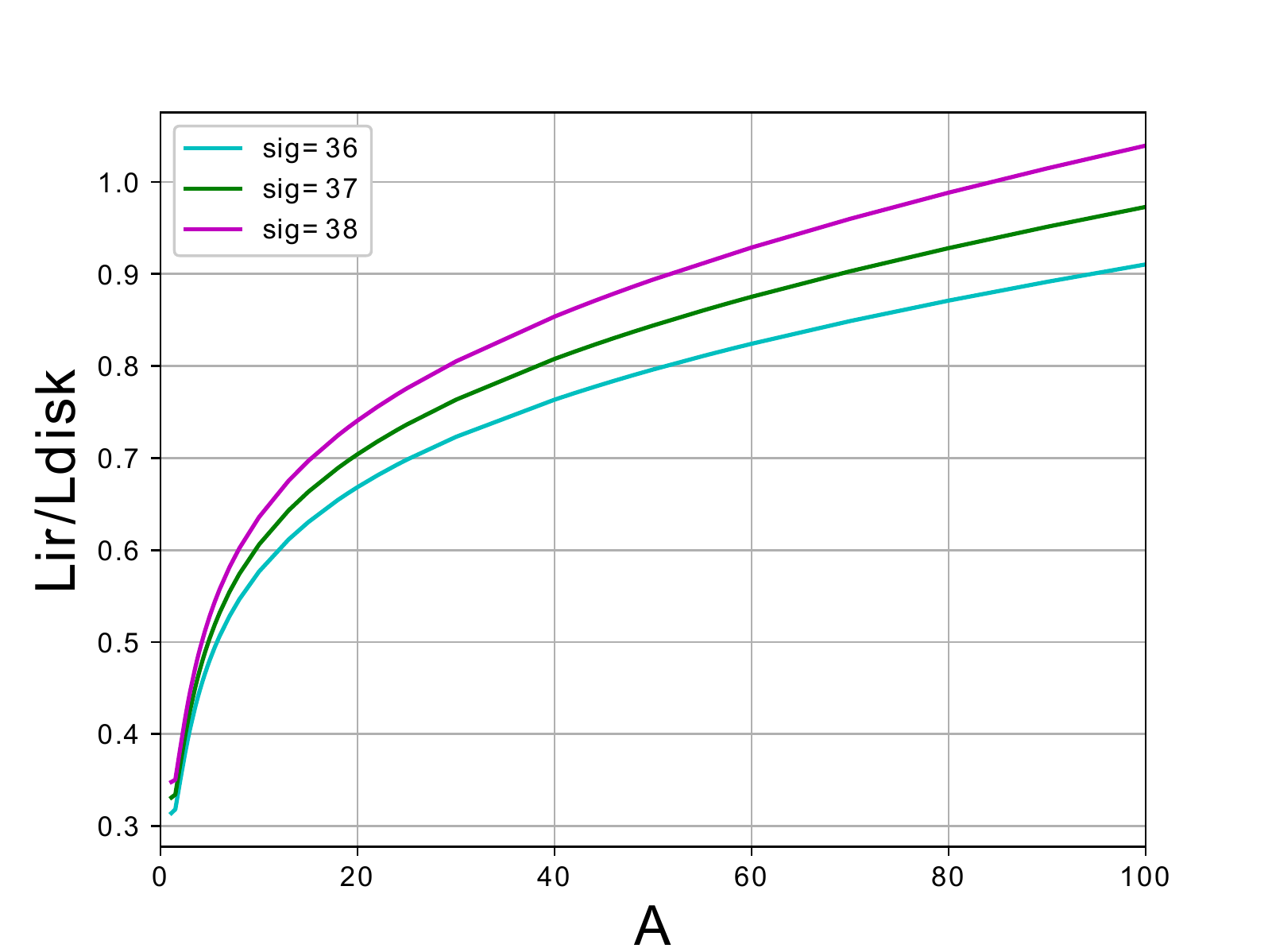}
    \caption{The illustration of the IR emission dominance with the increase of the torus optical depth for three values of the torus opening angles.}
    \label{fig:IR_to_opt}
\end{figure}

The disk luminosity significantly depends both on the optical depth along the equatorial plane ('A'), and on the width or the opening ($\sigma$) of the torus. Thus, with a single constraint from the data at a given redshift, we obtain a relation between the two model parameters, separately for each redshift. 
In Figure~\ref{fig:general}, we have considered the entire possible range of $\sigma$ between 0$^{\circ}$ (at the equatorial plane) to 90$^{\circ}$ (at the symmetry axis). The red curve shows the range of $\sigma$ for low redshift sources (Table~\ref{tab:redshift} for CF = 0.478;  \citealt{Gu_2013}) and green curve covers the $\sigma$ for high redshift sources (Table~\ref{tab:redshift}, CF = 0.871).
For the low redshift sources the maximum $\sigma$ allowed is 45$^{\circ}$, shown as an upper dotted blue horizontal line in Figure~\ref{fig:general}. This upper-limit comes from the fact that the average viewing angle for type 1 sources is $\sim$40$^{\circ}$ (\citealt{Czerny_2011}), which allows the maximum $\sigma$ to go close to 50$^{\circ}$. Hence to be on safe side, we have constrained our analysis to $\sigma$ equals to 45$^{\circ}$  as an upper limit.
The lower dotted blue horizontal line is constrained by the minimum allowed value of $\sigma$ for high redshift sources in our sample. More details about constraining the $\tau_{equatorial}$ are discussed in the section 3.2.
The two lines never cross which means that we do not see any possibility to have the same torus mean parameters for low and high redshift sources. However, we have the region where the $\sigma$ is common for both low and high redshift sources, that means that on average we can have similar opening angle of dusty torus in SDSS DR9Q and DR7Q catalog(\citealt{Gu_2013}), i.e.  at redshifts $\sim$ 0.8 and $\sim$ 2.2. However, even in this case the mean effective viewing angle is not identical, since it also depends on the extinction which is set by the parameter A equal to $\tau_{equatorial}$. We illustrate it in Figure~\ref{fig:i_eff_fixed_sigma}. 
We choose to keep the maximum value of $\tau_{equatorial}$ $\sim$ 100, which has also been suggested by \citet{Nenkova_2008} for a single cloud in their clumpy torus model.
Of course, also the observational appearance of the source is strongly affected since the ratio of the IR emission to the optical/UV emission also steeply rises with the torus optical depth (see Figure~\ref{fig:IR_to_opt}). The CF increases very sharply below A = 10, and after that the rise slows down and CF finally almost saturates for higher value of 'A'. The parameter 'A' can have large range of values starting from 1 (constrained by equation (10)) to any-other higher value. We have discussed the possibility of constraining the upper limit of 'A' in the section 3.2.

\subsection{\bf Available parameter space} \label{subsec:hide}


To constrain the upper limit of 'A', we consider the case of Compton thick type AGN. The maximum value of the torus optical depth is not easy to measure, \citet{Burlon_2011}  gives the constraint for the column density, N$_H$ $>$ 10$^{25}$ cm$^{-2}$. Now to calculate the optical depth along the equatorial plane (A or $\tau_{equatorial}$), we assume N$_H$ = 5$\times$10$^{25}$ cm$^{-2}$, and hence $\tau_{equatorial}$ $\sim$ $\sigma_T$*N$_H$, gives $\tau_{equatorial}$ $\sim$ 33, where $\sigma_T$ is the Thompson scattering cross section.

We adopt this value as the upper limit, and therefore Figure~\ref{fig:general} shows the plot for $\tau_{equatorial}$ and $\sigma$ for low ($z\sim$ 0.8) and high ($z\sim$ 2.2) redshift sources, where A is constrained between 1--33 and $\sigma$ between 25$^{\circ}$ -- 45$^{\circ}$ and 38$^{\circ}$ -- 85$^{\circ}$ for low and high redshift sources, respectively. 


\begin{figure}
\centering
\includegraphics[scale=0.45]{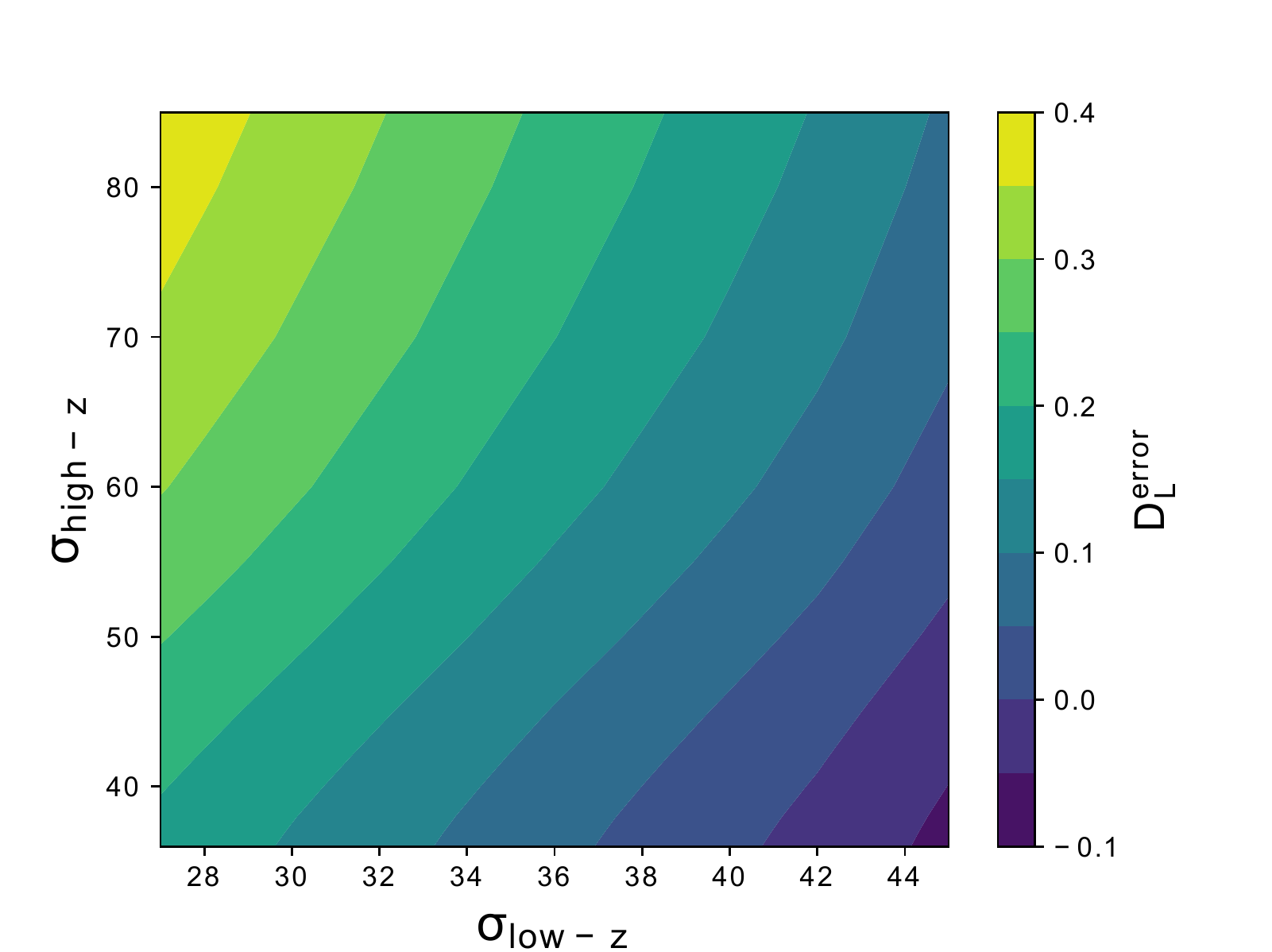}
\caption{The contour plot for the error on luminosity distance estimated from the luminosity distance ratio of low and high redshift sources.}
\label{fig:DL}
\end{figure}

\begin{figure}
    \centering
    \includegraphics[scale=0.45]{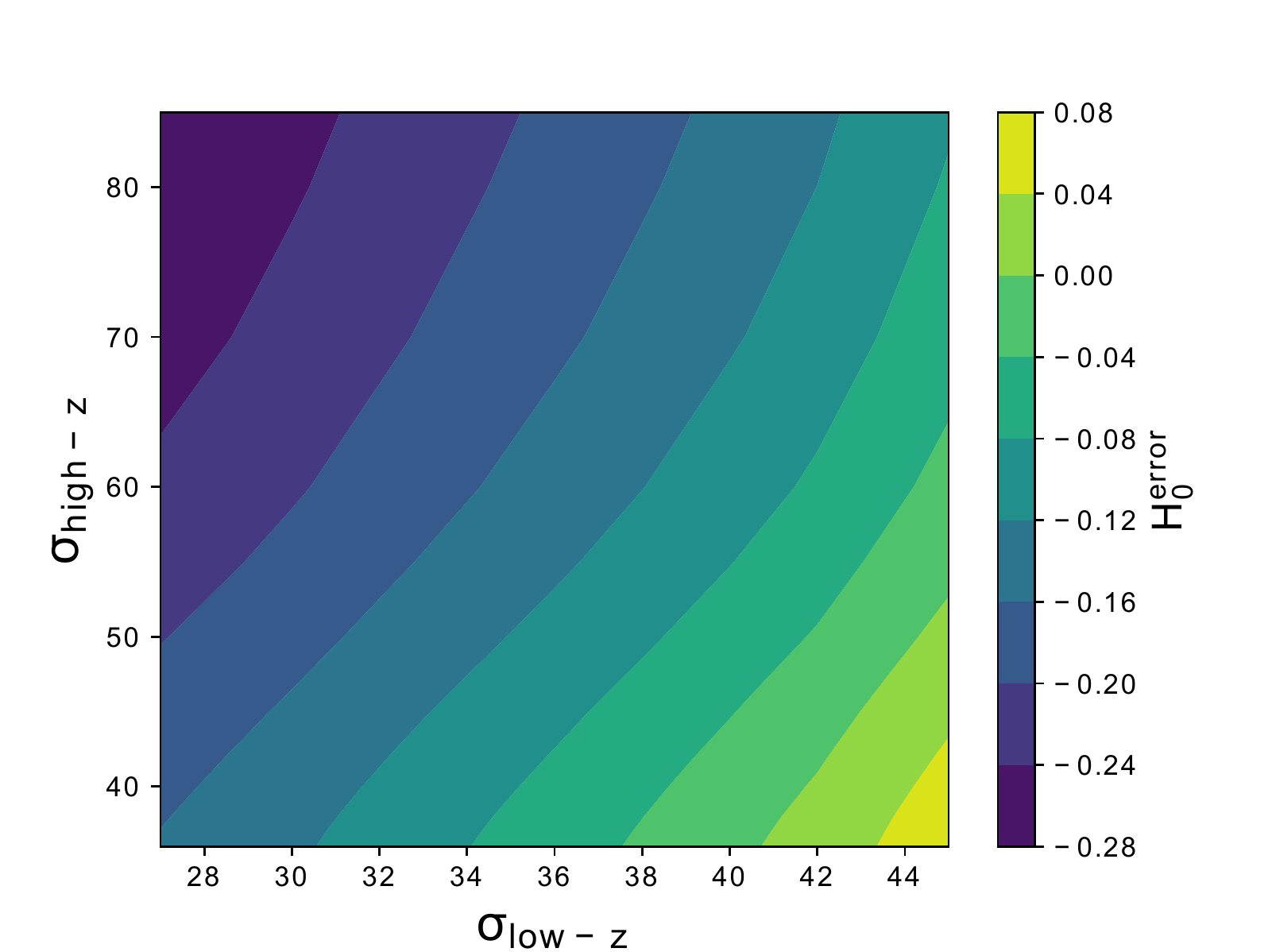}
    \caption{The contour plot for the error on H$_0$ estimated from the equation (22).}
    \label{fig:H0_error}
\end{figure}

\subsection{\bf Effect on the Luminosity distance}

Having constrained the $\sigma$ for low and high redshift sources from \citet{Gu_2013}, we can now estimate the potential error on the determination of the luminosity distance, if the effect of the change of geometry between the low and high redshift is neglected, making use of Equation~\ref{eq:DL_rat}.

We pick up the torus parameters from the two lines shown in Figure~\ref{fig:general}, which forms a 2-D plot of the possible errors in the luminosity distance, parameterized by the values of $\sigma$ on the low redshift branch and at high redshift branch. 
The possible errors on luminosity distances at higher redshift due to viewing angle and the torus extinction are plotted in Figure~\ref{fig:DL}. The
color bar of the contour plot shows the relative error of the luminosity distance, and X $\&$ Y-axis represent the half opening angle at lower and higher redshifts respectively. At the higher redshift, if the half opening angle of the torus is less than 45$^{\circ}$ (i.e. $\sigma$ $<$ 45$^{\circ}$), we underestimate the luminosity distance. However, if $\sigma$ $>$ 45$^{\circ}$, we overestimate the luminosity distance. The error on luminosity distance for maximum possible $\sigma$ would be $\sim$40 $\%$ but this would happen only if the true torus opening angle at high redshift is as high as 80$^{\circ}$. If the used parameter space is too broad, and in reality the torus at high redshift has the opening angle about 40$^{\circ}$, then the maximum possible error is smaller, below 20 $\%$. This error is just an approximation and it can be modified in the future by considering more appropriate model of torus geometry and measuring the viewing angle of the individual objects, which will constrain much better the parameter space. 


\subsection{\bf Potential error for Cosmological parameters}

Since the luminosity distance is used to measure the Hubble constant in local universe and to determine other parameters of the cosmological model, we see that potential negligence of the systematic trend in the viewing angle of the torus may lead to considerable bias in determining these parameters from quasar reverberation studies. To see how large are potential errors in quasar-based method we translate the error in determination of the luminosity distance to an error in cosmological parameters. With two-redshift approach based on the data from \citet{Gu_2013} and our model we can only provide a simple illustration.

First we assume that standard $\Lambda$CDM model describes correctly the expansion of the Universe and that we know the values of the parameters $\Omega_{\Lambda}$ and $\Omega_m$, but we aimed at determination of the Hubble constant. Using Equation~\ref{eq:H0_incorrect} we thus determine the error in $H_0$ resulting from the incorrect representation of the quasar geometry at high redshift of 2.2. The result is shown in Figure~\ref{fig:H0_error}. The contour plot has a similar shape to the error in luminosity distance but relative errors are much smaller. The errors in a whole parameter space are up to 18 \%, but even if the opening angle of the torus at high redshift is actually of order of 40$^{\circ}$, the error could reach 16 \%. Since the error can be both positive and negative, with luck we can have no error. However, if the aim of the quasar-based method is to meet the limits of 1  \%, achieved with SN Ia \citep{riess2020}, then the issue of the viewing angle trend in AGN cannot be ignored. From our plot we can estimate that in order to have 1 \% accuracy in $H_0$ we would need to determine (statistically) the values of $\sigma$ with 0.5$^{\circ}$ accuracy. 

\begin{figure}
    \centering
    \includegraphics[scale=0.45]{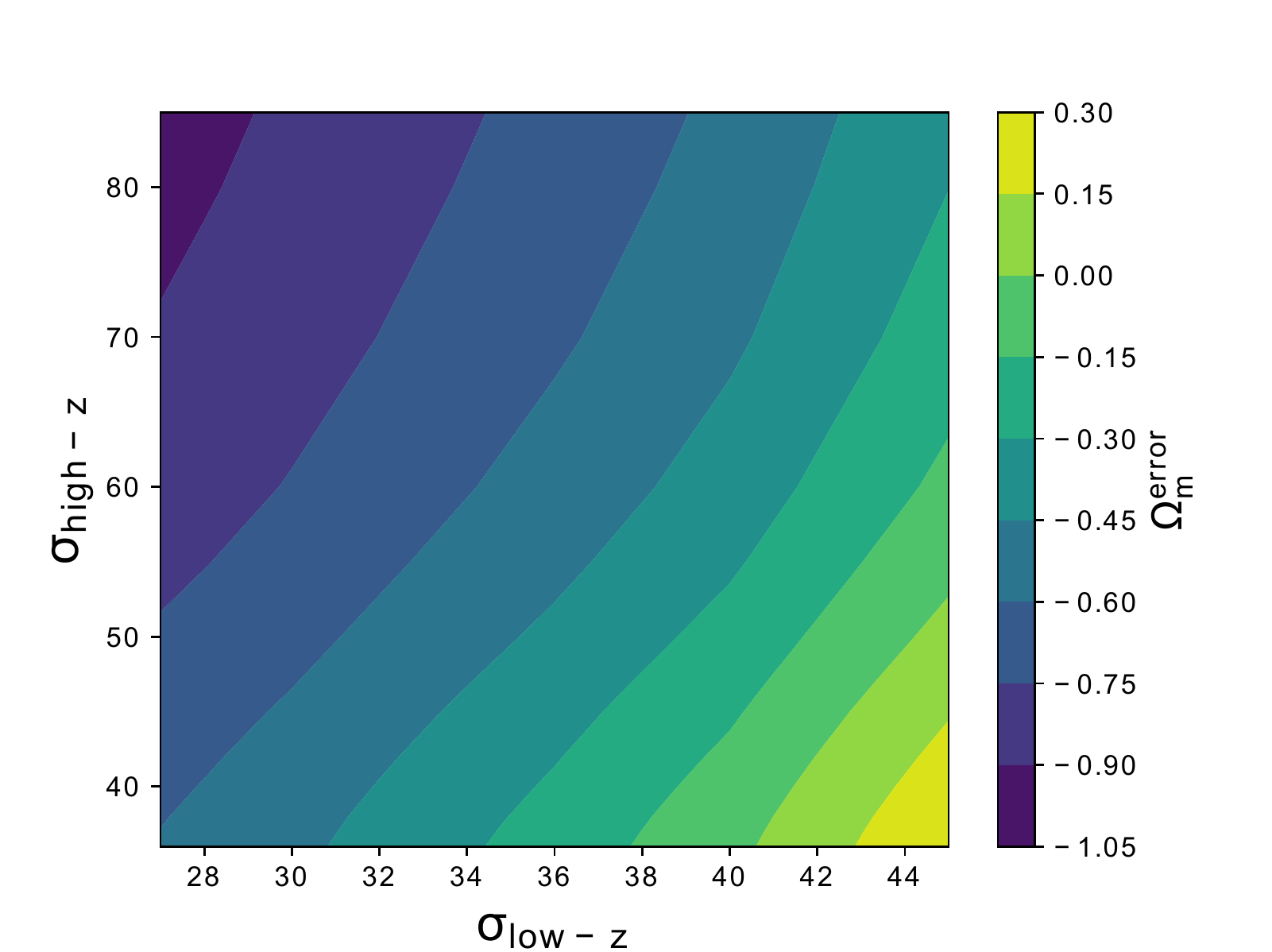}
    \caption{The contour plot for the error on $\Omega_m$ estimated from the equation (23).}
    \label{fig:Omega_m_error}
\end{figure}

Next we assume that we know the Hubble constant, but we wish to use quasars to estimate the other cosmological parameters, for example $\Omega_m$, assuming that the Universe is flat. Using now Equation~\ref{eq:Omega_m_incorrect} we derive the contour error map for the $\Omega_m$ (see Figure~\ref{fig:Omega_m_error}). We see that both negative and positive errors are expected, and if the broad parameter range used for the plot is allowed the error is huge. If the realistic values of $\sigma_{high-z}$ are rather close to 40$^{\circ}$, the error is smaller, but achieving an error of order of 1\%, required for precision cosmology, will clearly require taking the systematic errors connected with the viewing angle into account.

We do not claim here that the systematic errors in the published papers are actually as large as illustrated in Figures \ref{fig:DL} -- \ref{fig:Omega_m_error} since we use here only one-parameter data input and 2-parameter models so some ranges of the parameter space are unlikely. However, we expect that the plots well argue that the effect is important and requires further studies to validate cosmological applications. In principle, we could use exactly the approach of \citet{Gu_2013} and interpret directly the measured ratio as the viewing angle, thus assuming that the torus can be modeled as a solid body, without extinction or disk anisotropy included. In that case, we would obtain a unique result from their measurements: the viewing angle in the lower redshift sources would be 40.5$^{\circ}$, and at high redshift sample 55.6$^{\circ}$. Under those strong assumptions, the corresponding error in the luminosity distance would then be 16\%. 


\section{\bf Discussions}
The structure of AGNs can be divided in two parts. The central part (BH + disk + BLR) of bright AGN is dominated by the optical and UV emission from the flat optically thick disk.  At larger distance the central parts are partially surrounded by the dusty torus, which emits mostly the reprocessed emission in the infrared band. The dusty torus can be optically thin or optically thick. The emission from the central part of the AGN is anisotropic because the flat disk emission is anisotropic, and the optically thick dusty torus further modifies the radiation in an angle-dependent way. Therefore, the observed bolometric luminosity of the AGNs depends on the viewing angle. The issue is well known, \citet{Netzer_1987} discussed  how the central radiation source in AGN was responsible for the anisotropic emission, but it was not taken into account in the context of quasar applications to cosmology.

In this paper we used the observational data from \citet{Gu_2013} and a very simple torus model to evaluate implications of this effect when quasars are used for cosmology. Our toy model of torus is continuous, with optical depth dependent on the angular distance from the equatorial plane, we assumed that the disk emission (also dependent on the viewing angle) is filtered by such a torus, and we obtained constraints for the torus structure at two values of redshift corresponding to the mean properties of the sample from \citet{Gu_2013}. Later we assumed that the cosmological method is calibrated at lower redshift and we checked what happens if the change of geometry is ignored when the study of high redshift sample is done. We specifically had in mind the reverberation method when the absolute luminosity of a quasar is determined either from the time delay of the BLR lines with respect to the continuum, or from the relative delays of two continuum bands. Such time delay depends mostly on the actual size of the source, and only weakly on the source inclination while the observed flux does depend on inclination and obscuration.

Considering the effect of viewing angle and the torus optical depth, we have estimated the bolometric luminosities and the luminosity distance for high redshift AGNs, applying either correct geometry, or applying the geometry appropriate for the low redshift quasar sample. We then calculated the error that one would make on measurement of luminosity distance if one does not consider the effect of the change in the viewing angle and torus optical depth (Figure~\ref{fig:DL}). The result shows that the errors are quite significant and they can not be ignored. 
Since the luminosity distance can be used to do the cosmology, the error on cosmological model can also be estimated. The errors on H$_0$ and $\Omega_m$ are shown in Figure~\ref{fig:H0_error} $\&$ \ref{fig:Omega_m_error}.

For a single source, as could be seen from Figure~\ref{fig:H0_error}, the 4$\%$ error in H$_0$ is equivalent to the difference in the $\sigma$ by $\sim$ 2$^{\circ}$. We know that $\sigma$ is directly related to the viewing angle $i$ and hence the change in $i$ would be order of 2$^{\circ}$. As it has been seen in many studies \citep{riess2020} that the 1$\%$ error in H$_0$ is significant enough to formally claim the tension in H$_0$. And hence from a single object to claim the error of 1$\%$ in H$_0$ we need the accuracy of 0.5$^{\circ}$ in effective viewing angle.  The viewing angle estimated in a single source with such an accuracy would be rather unrealistic but for cosmological applications we need only to know the trend in the statistical sense. Thus, if in a single source we can have an estimate of the viewing angle with 10$^{\circ}$ accuracy, we would need 400 sources in a redshift bin to have statistically the requested accuracy.

{\bf \subsection{The accretion disk model}}

\begin{figure}
    \centering
    \includegraphics[scale=0.45]{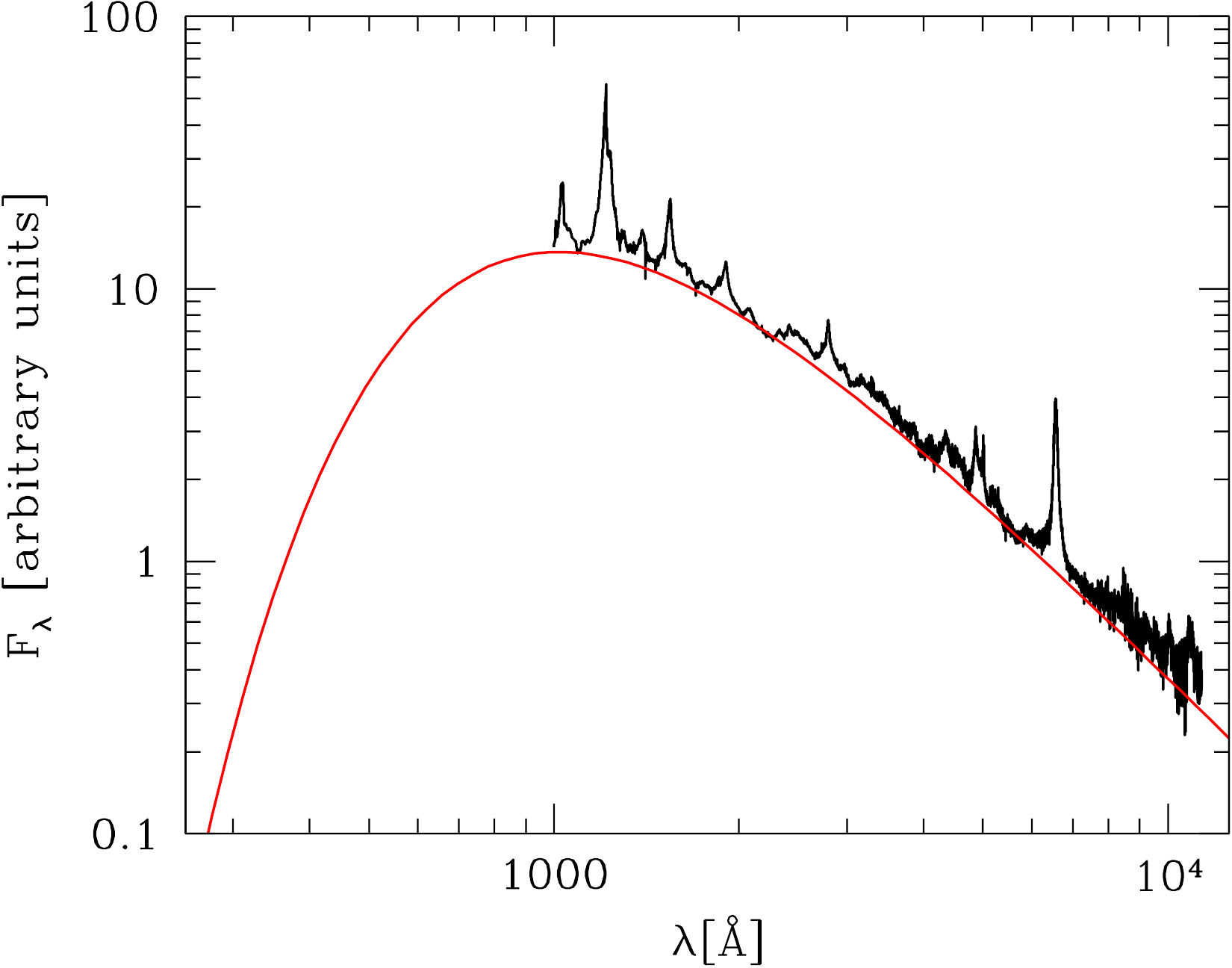}
    \caption{The composite spectrum from \citet{selsing2016} (black line) compared to the standard \citet{ss1973} accretion disk model (red line) for parameters: black hole mass $6 \times 10^8 M_{\odot}$, and the Eddington ratio  $L/L_{Edd}$ = 0.3.}
    \label{fig:xshooter}
\end{figure}

In the current paper we used the simplest possible representation of the accretion disk (see Equation~\ref{eq:disk}) without using explicitly the disk spectral shape. The important fact we needed was the strict proportionality of the observed flux to $\cos i$, which is correct for the standard Shakura-Sunyaev disk model \citep{ss1973}. This simple model applies quite well to bright quasars, in the optical and near-UV range. As an illustration, in Figure~\ref{fig:xshooter} we show such a model fitted to the recent quasar composite spectrum based on XSHOOTER data \citep{selsing2016}.  The model represents well the optical data, although at the longest wavelengths traces of the host galaxy contribution start to be seen. However, at far-UV and in the X-ray band, the standard model does not apply, and the soft and hard X-ray emission come from the innermost part of the disk, from the warm and hot corona. In the IR, the torus emission and the contribution from the host galaxy dominate the broad band spectrum (for observational discussion of broad band quasar spectra see \citealt{richards2006}). Modelling this broad band emission is usually done using some parametric models, usually aimed at specific energy band. Modelling IR emission requires a model of the dusty torus, modelling the X-ray spectra requires a model of the corona/coronas and the model of the effects of reprocessing. For just the broad-band spectrum, the convenient model with two coronas was proposed by \citet{kubota2018}. The theoretical modelling of the innermost part of the disk is complex (see e.g. \citealt{czerny_rev_2018} for a recent review), particularly if the important role of the magnetic field is postulated. At very high accretion rates additionally the advection has to be included \citep{abramowicz1988}. At much lower Eddington rates than typically seen in quasars an inner hot flow replaces the cold disk \citep[e.g.][]{ichimaru1977,adaf1994,melia2001}. These processes do not affect the optical emission considerably. However, if the broad-band spectrum is used, then some representation of these processes is required.\\

{\bf \subsection{Advanced torus models}}
There are various methods to estimate viewing angles of quasars. First of all, more advanced torus models combined with the good SED data can be used for that purpose. In past, various models have been proposed in the context of dusty torus based on the X-ray observations. The model \textit{MYtorus} \citep{Murphy_2009} and \textit{BNtorus} \citep{Brightman_2011} have been used extensively for the spectroscopic studies of the obscured AGNs. However, they have very limited number of parameters to describe the physical scenarios of torus. In \textit{MYtorus} model the covering factor is fixed to 50$\%$. In \textit{BNtorus} model the opening angle of the torus is considered as a free parameter but the column density along the line of sight is assumed to be equal to the torus column density (which is the case for Compton thick AGN). The torus model used in our study is more like BNtorus model where the opening angle varies from 0$^{\circ}$ -- 90$^{\circ}$ (along the equatorial plane -- symmetric axis) and the column density we have chosen is more like Compton thick AGN (N$_H$ $\sim$ 10$^{25}$ cm$^{-2}$). 
The recent dusty torus model (borus02) is proposed by \citet{Balokovic_2018} where they assumed the toroidal geometry similar to the BNtorus model but the new model is more flexible in terms of available parameters space, it allows for a different line of sight column density of the torus than the total torus column density. Also the covering factor is a free parameters and it varies with torus opening angle (CF = cos $\theta_{torus}$) measured from the symmetry axis. They also assumed the smooth-density torus scenario which reasonably represents the physical reality of torus. Along with the smooth density torus there are studies which shows the possibility of clumpy torus also. A recent study by \citet{Buchner_2019} discussed the clumpy torus model based on the X-ray studies. The clumpiness in the torus has been observationally confirmed through the X-ray eclipse events as reported by \citet{Risaliti_2002} $\&$ \citet{Markowitz_2014}. 
The \textit{XCLUMPY} model of \citet{Buchner_2019} explains such events. 

A number of other interesting studies were also done addressing the aspect of the quasar radiation anisotropy. \citet{venanzi2020} has studied the impact of anisotropic radiation from central AGN on the emergence of dusty winds. Their simulation shows that the anisotropic radiation from central source changes the outflow opening angle by making the wider cone with respect to isotropic radiation. Our model also assumed the anisotropic AGN radiation where more radiation is expected along the symmetric axis than towards the dusty torus. \citet{yamada2020} in their studies included not only the effect of the torus but also the dusty polar outflows in their study of anisotropy. The effect of orientation of individual AGN on the dispersion on their luminosity and line width was included by \citet{negrete2018} in their study based on the virial broadening estimator, where they also included the limb darkening effect of the disk emission, neglected in our approach. However, if even a small fraction of the central flux is back-scattered towards the disk, the limb darkening change to limb brightening effect \citep[see e.g.][and the references therein]{zycki1994}, and any enhancement of the dissipation profile within the disk close to the surface will also give the same effect \citep[see e.g.][]{rozanska2015,petrucci2020}.

Recent study by \citet{bisogni2019} shows that the low EW[O III] sources are face-on view and have more flatter IR/optical SED compared to the high EW[O III] sources, which have edge-on view with respect to the observer. They also claim that their finding suggest the torus to be clumpy.\\
 Although the clumpy torus model is the best we have currently, a more advanced torus model which can measure the torus parameters and the viewing angle independently, is needed in the future.\\

{\bf \subsection{Viewing angle as a solution to the quasar-based claims of tension in cosmology}}

In our paper we mostly focused on the future corrections to quasar-based cosmological methods using the time delay measurements. However, the issue of the potential problem also applies to the most developed quasar method proposed by \citet{risaliti2015}. They have tested the $\Lambda CDM$ cosmological model by using the nonlinear relation between the UV and X-ray luminosities of quasars. 
In the recent paper by \citet{Risaliti_2019}, they have considered the quasars in the redshift range from 0.5 upto 5.5 and plotted the Hubble diagram. Their results shows that the value of Hubble parameter H$_0$ matches with \citet{Suzuki_2012} at lower redshift (0-1.4). However, at higher redshift it deviates from the $\Lambda$CDM model at statistical significance of 4$\sigma$ (\citealt{Risaliti_2019}). Considering the effect of viewing angle in their equation (2) \citep{risaliti2015}, we derive the luminosity distance dependence on viewing angle, i.e. D$_L$ $\sim$ (cos$i$)$^{0.75}$ for $\gamma$ $\sim$ 0.6, even steeper than D$_L$ $\sim$ (cos$i$)$^{0.5}$ characteristic for reverberation-based methods. 

\begin{figure}
    \centering
    \includegraphics[scale=0.45]{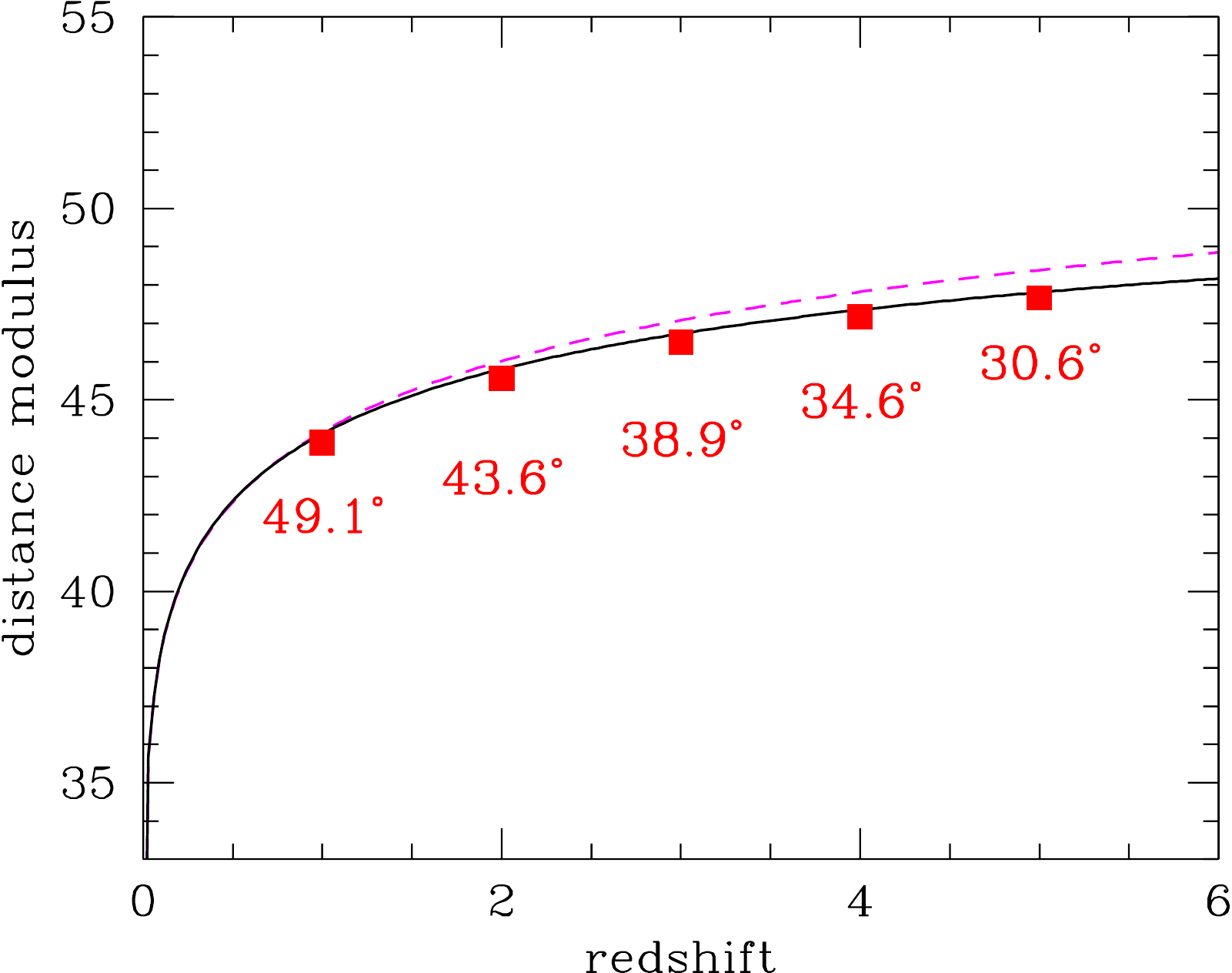}
    \caption{The distance modulus in the standard $\Lambda$CDM model (dashed magenta line, the best fit to the SNIa ans quasar data from \citet{Risaliti_2019} (black continuous line), and our points (in red) representing the quasars for a standard model but with the viewing angle included (model with $\delta = 0.25$, and $cos i_0 = 0.55$). Corresponding values of the mean viewing angle are marked on the plot.}
    \label{fig:Risaliti}
\end{figure}

Assuming that the extinction is unimportant, we illustrate how the viewing angle can mimic the departure from the standard ($\Lambda$CDM) model. We assume that, statistically, the effective viewing angle $i$ of AGN may vary as $cos i = cos i_0(1 + z)^{\beta}$. We plot the distance modulus for the standard $\Lambda$CDM model in Figure~\ref{fig:Risaliti}. We also plot the best fit to the quasar data provided by \citet{Risaliti_2019} in the form of a cosmographic approach, taking their values of the best fit parameters. We should note here that the use of this expansion is questioned by \citet{banerjee2020}, but it represents the observational results from Figure 2 of  \citet{Risaliti_2019}. If we fix the slope $\beta = 0.3$, our points roughly reproduce the observed departure. The requested change of the viewing angle (values provided in the plot) is considerable, but plausible. If the extinction plays a role like in our torus model (see Section~\ref{sect:model}), the requested change of the viewing angle might be even lower.

Our example does not imply that only the standard $\Lambda$CDM is correct and the tension must be attributed to the viewing angle issues. The tension can be real, particularly if the claim is based on other probes than quasars. The need for a revision of the standard model is now proposed in numerous papers (\citealt{Wong_2020}, \citealt{Riess_2019}, \citealt{lusso2019}) and
there are already some studies that have discussed e.g. the possibility of R$_{\rm h}$ = $ct$ universe instead of the $\Lambda CDM$ model to explain the Hubble tension based on Quasar data (\citealt{Melia_2019}, \citealt{2016IJMPD..2550060L}). It is also possible that the claim of the departure from the standard model in \citet{Risaliti_2019} is not justified; \citet{Melia_2019} pointed out that the justification for introducing more free parameters (1 for standard $\Lambda$CDM and 2 for more general expansion) have to be done properly, and some aspects of the analysis itself were questioned by \citet{yang_tao2019}.
However, in this paper, we only concentrate on demonstrating that if quasars are used for cosmology, and the method involves the use of the quasar observed flux in the optical/UV band, the viewing angle issue needs to be addressed.


Thus the future quasar-based studies aimed to do the precision cosmology should carefully address the issue of the potential statistical evolution of the mean torus properties with the redshift.
In their paper \citet{khadka2020} considered a combination of different cosmological probes and concluded that the observed disagreement with $\Lambda$CDM model at redshifts from 2 to 5 is most likely caused by problems with QSO data. That seems like an important warning.\\

\section{Summary}

Our work is based on the results from \citet{Gu_2013} which show the redshift evolution of CF between two samples of quasars at two values of redshift: $\sim 0.8$ and $\sim 2.2$. The higher covering factor at higher redshift suggests that the structure and density distribution of dusty torus could be different than at the low redshift quasars. This change would affect statistically the viewing angles of the sources and their observed luminosity.   
This work is a pilot study to show how the viewing angle can play an important role in measuring the luminosity distance and further the cosmological parameters. We have developed an analytical method to estimate the CF by considering the viewing angle and smooth density distribution of dusty torus. Using our method we have estimated the possible error on the luminosity distance that we could make if we do not include the viewing angle. 
Our results in Figure~\ref{fig:DL} show the error on luminosity distance might go up to 40 $\%$ for high CF sources (with high torus opening angle). This effect should be verified by measuring the large sample of quasars, with much better modelling of dusty torus reprocessing and much better data coverage. Our study shows that to have 1$\%$ error in H$_0$ one need to have a statistical accuracy of 0.5$^{\circ}$ in effective viewing angle in the quasar sample, and for that one will need nearly 400 sources in a single redshift bin. 




\acknowledgments
We thanks anonymous referee for valuable comments that helped us to improve the manuscript.
The project was partially supported by the Polish Funding Agency National Science Centre, project 2017/26/A/ST9/00756 (MAESTRO 9), and MNiSW grant DIR/WK/2018/12. 

%

\vspace{5mm}




\bibliography{sample63}{}
\bibliographystyle{aasjournal}

\end{document}